\documentclass[letter,12pt, epsfig]{article}

\usepackage{graphicx, psfrag}

\usepackage{epsfig}
\usepackage{amssymb}
\usepackage{amsfonts}
\usepackage{amsmath}

\usepackage{graphicx}
\usepackage{caption}
\usepackage{subcaption}

\usepackage{tikz}
\usepackage{epstopdf}
\usetikzlibrary{backgrounds, arrows,positioning,shapes,decorations.markings,snakes}

\newskip\humongous \humongous=0pt plus 1000pt minus 1000pt

\newif\ifdtup

\catcode`\@=11

\@addtoreset{equation}{section}
\def\theequation{\thesection.\arabic{equation}}

\def\@normalsize{\@setsize\normalsize{15pt}\xiipt\@xiipt
\abovedisplayskip 14pt plus3pt minus3pt%
\belowdisplayskip \abovedisplayskip
\abovedisplayshortskip \z@ plus3pt%
\belowdisplayshortskip 7pt plus3.5pt minus0pt}

\def\small{\@setsize\small{13.6pt}\xipt\@xipt
\abovedisplayskip 13pt plus3pt minus3pt%
\belowdisplayskip \abovedisplayskip
\abovedisplayshortskip \z@ plus3pt%
\belowdisplayshortskip 7pt plus3.5pt minus0pt
\def\@listi{\parsep 4.5pt plus 2pt minus 1pt
     \itemsep \parsep
     \topsep 9pt plus 3pt minus 3pt}}

\relax

\catcode`@=12

\catcode`\@=11

\def\section{\@startsection{section}{1}{\z@}{3.5ex plus 1ex minus
   .2ex}{2.3ex plus .2ex}{\large\bf}}

\def\thesection{\arabic{section}}
\def\thesubsection{\arabic{section}.\arabic{subsection}}

\def\appendix{\setcounter{section}{0}
 \def\thesection{Appendix \Alph{section}}
 \def\thesubsection{\Alph{section}.\arabic{subsection}}
 \def\theequation{\Alph{section}.\arabic{equation}}}

\def\SymBoxes#1#2#3#4{\newdimen\un@t \un@t#3%
\raisebox{#1}{\rule{#2\un@t}{#4}\hskip-#2\un@t
\@tempdimb\un@t \advance\@tempdimb by-#4\@tempcntb#2\relax%
\@whilenum{\@tempcntb>0}\do{
\rule{#4}{\un@t}\hskip\@tempdimb \advance\@tempcntb by\m@ne}%
\hskip-#2\un@t \rule[\un@t]{#2\un@t}{#4}%
\rule[\un@t]{#4}{#4}\hskip-#4
\rule{#4}{\un@t}}\hskip-#4}                

\begin{document}

\newcommand{\beq}{\begin{equation}}
\newcommand{\eeq}{\end{equation}}
\newcommand{\bea}{\begin{eqnarray}}
\newcommand{\eea}{\end{eqnarray}}
\newcommand{\beas}{\begin{eqnarray*}}
\newcommand{\eeas}{\end{eqnarray*}}
\newcommand{\defi}{\stackrel{\rm def}{=}}
\newcommand{\non}{\nonumber}
\newcommand{\bquo}{\begin{quote}}
\newcommand{\enqu}{\end{quote}}
\renewcommand{\(}{\begin{equation}}
\renewcommand{\)}{\end{equation}}
\def\IZ{{\mathbb Z}}
\def\IR{{\mathbb R}}
\def\IC{{\mathbb C}}
\def\IQ{{\mathbb Q}}

\def\g{\gamma}
\def\m{\mu}
\def\n{\nu}
\def\a{\alpha}
\def\b{\beta}

\def\CM{{\mathcal{M}}}
\def\dCM{{\left \vert\mathcal{M}\right\vert}}

\def \d{\textrm{d}}
\def \p{\partial}

\def \Pf{\rm Pf\ }

\def \pr{\prime}

\def\Tr{ \hbox{\rm Tr}}
\def\half{\frac{1}{2}}

\def \eqn#1#2{\begin{equation}#2\label{#1}\end{equation}}
\def\de{\partial}
\def\Tr{ \hbox{\rm Tr}}
\def\H{ \hbox{\rm H}}
\def\HE{ \hbox{$\rm H^{even}$}}
\def\HO{ \hbox{$\rm H^{odd}$}}
\def\K{ \hbox{\rm K}}
\def\Im{ \hbox{\rm Im}}
\def\Ker{ \hbox{\rm Ker}}
\def\const{\hbox {\rm const.}}
\def\o{\over}
\def\im{\hbox{\rm Im}}
\def\re{\hbox{\rm Re}}
\def\bra{\langle}\def\ket{\rangle}
\def\Arg{\hbox {\rm Arg}}
\def\Re{\hbox {\rm Re}}
\def\Im{\hbox {\rm Im}}
\def\exo{\hbox {\rm exp}}
\def\diag{\hbox{\rm diag}}
\def\longvert{{\rule[-2mm]{0.1mm}{7mm}}\,}
\def\a{{\textsl a}}
\def\dag{{}^{\dagger}}
\def\tq{{\widetilde q}}
\def\p{{}^{\prime}}
\def\W{W}
\def\N{{\cal N}}
\def\hsp{,\hspace{.7cm}}
\newcommand{\C}{\ensuremath{\mathbb C}}
\newcommand{\Z}{\ensuremath{\mathbb Z}}
\newcommand{\R}{\ensuremath{\mathbb R}}
\newcommand{\rp}{\ensuremath{\mathbb {RP}}}
\newcommand{\cp}{\ensuremath{\mathbb {CP}}}
\newcommand{\vac}{\ensuremath{|0\rangle}}
\newcommand{\vact}{\ensuremath{|00\rangle}}
\newcommand{\oc}{\ensuremath{\overline{c}}}
\newcommand{\IP}{\ensuremath{\mathbb P}}
\newcommand{\A}{\ensuremath{\mathbb A}}
\newcommand{\T}{\ensuremath{\mathbb T}}

\def\M{\mathcal{M}}
\def\F{\mathcal{F}}
\def\d{\textrm{d}}

\def\eps{\epsilon}

\begin{flushright}
\end{flushright}
\thispagestyle{empty}

\vspace{18pt}
\begin{center}
{\Large \textbf{On the four-point function of the stress-energy tensors in a CFT}}
\end{center}

\vspace{6pt}
\begin{center}
{\large\textsl{Anatoly Dymarsky }\\}
\vspace{25pt}
\textit{\small Department of Applied and Theoretical Physics, \\University of Cambridge, Cambridge, UK}\\ \vspace{6pt}

\end{center}

\vspace{12pt}

\begin{abstract}
We discuss to what extent the full set of Ward Identities constrain the four-point function of the stress-energy tensors or conserved currents in a conformal field theory. We calculate the number of kinematically unrestricted functional degrees of freedom governing the corresponding correlators and find that it matches the number of functional degrees of freedom governing scattering amplitudes of some ``dual'' massless particles in the auxiliary Minkowski space. We also formulate the conformal bootstrap constraints for the correlators in question in terms of only unrestricted degrees of freedom. As a by-product we find interesting parallels between solving Ward Identities in coordinate and momentum space.
\end{abstract}
\titlepage

\section{Introduction}
Conformal symmetry imposes powerful constrains on the correlation functions of primary operators in a conformal field theory. In particular, all two and three-point functions are fixed up to a small number of normalization constants \cite{Polyakov:1970xd, Polyakov:1974gs, Osborn1, Osborn2}.  At the same time the form of the four-point functions is not universal and depends on functions which encode the CFT's dynamics. In case of four scalars there is just one such function of the cross-ratios $u,v$
\bea
\label{4pt}
\langle {\mathcal O}(x_1) {\mathcal O}(x_2) {\mathcal O}(x_3) {\mathcal O}(x_4)\rangle={f(u,v)\over (x_1-x_2)^{2\Delta}(x_3-x_4)^{2\Delta}}\ ,\\ \nonumber
u={(x_1-x_2)^2(x_3-x_4)^2\over (x_1-x_3)^2(x_2-x_4)^2}\ ,\quad
v={(x_1-x_4)^2(x_2-x_3)^2\over (x_1-x_3)^2(x_2-x_4)^2}\ .
\eea 
When the operators have spin ${\mathcal O}_{\mu\dots}$ the four-point correlator depends on many such functions $f^I(u,v)$.

In a general case the expression \eqref{4pt}, or its generalization when the  operators have spin, solves all Ward Identities\footnote{Namely, Poincar\'{e} invariance and covariance under special conformal transformations. Throughout the paper we keep all $x_i$ distinct and do not discuss those Ward Identities which involve coincident points.} with arbitrary $f^I(u,v)$. But in a special case when the operator dimension $\Delta$ saturates the unitary bound the operator ${\mathcal O}_{\mu\dots}$ becomes  conserved 
\bea
\partial_\mu {\mathcal O}_{\mu\dots}=0\ .
\eea
In this case there are additional Ward identities that require that  $\partial_\mu {\mathcal O}_{\mu\dots}$ inside any correlator vanishes. At the level of two-point function this condition follows from conformal algebra and is automatically satisfied. At the level of three-point function conservation provides a set of linear constraints on the normalization coefficients. In the case of four-point function conservation of $ {\mathcal O}_{\mu\dots}$ yields a set of first order differential equations on $f^I(u,v)$ which further restrict possible form of the correlation function in question.

In this paper we analyze this system of coupled equations and calculate the number of unconstrained  functional degrees of freedom governing the corresponding correlators. We were not able to solve these constrains explicitly. But we found that the number of the unconstrained degrees of freedom for the four-point function of the operators of spin $\ell_i$ in  $\rm d$ dimensions matches precisely the number of functions ${\sf f}^{\sf I}(s,t)$ of the Mandelstam variables  governing most general scattering amplitude of four particles of spin $\ell_i$ in the $({\rm d}+1)$-dimensional Minkowski space. Thus our findings support and generalize an interesting connection between the CFT correlators and scattering amplitudes first observed in  \cite{Hofman:2008ar} and further generalized in \cite{Costa:2011mg}.

Our motivation to identify the degrees of freedom unconstrained by the Ward Identities is rooted, besides the usual aspiration to solve all available kinematic constrains explicitly, in the desire to apply the conformal bootstrap approach to the correlators with spin.  The conformal bootstrap has proven to be a powerful tool to constrain CFT dynamics  in various dimensions \cite{CB}. Yet to this moment the applications were limited to the four-point functions of the identical scalar operators. Certainly, considering operators with spin should yield more information, but technically it is much more difficult as the number of cumbersome constraints grows rapidly with spin. A particularly interesting case would be to consider the four-point function of the stress-energy tensors because the stress-energy tensor is the most universal operator present in all CFTs. Hence one might expect the corresponding constraints to be most fundamental. Although the resulting number of constraints is large (in a general case there will be $633$ coupled equations), due to conservation $\partial_\mu T_{\mu\nu}=0$ many of them are not independent. If one succeeds to reformulate these constrains in terms of only unrestricted  degrees of freedom the number of equations would reduce drastically (e.g.~in $\rm d=3$ there will be just $5$ such equations). To develop such a formalism is one of the goals of this paper. 

This paper is organized as follows. In the next section we discuss general properties of the system of equations which encode conservation of operators at the level of correlation function. In particular we find the number of unrestricted degrees of freedom governing the four-point function of stress-energy tensors or conserved currents and propose the way to formulate the conformal bootstrap constraints without degeneracy. In section 3 we discuss  Ward Identities in the momentum space and establish an interesting parallel between imposing special conformal invariance in the coordinate space and conservation in the momentum space. In section 4 we calculate the number of functions governing scattering amplitudes of massless particles in an auxiliary Minkowski space and compare it with the number of unrestricted degrees of freedom calculated in section 2. We conclude with section 5.  

\section{Imposing Conservation in the Coordinate Space}
\label{sec:consconst}

In a conformal field theory the full set of Ward Identities can be understood in the following way. The underlying symmetries impose that the correlator of primary operators $\langle {\mathcal O^1}(x_1)\dots  {\mathcal O^n}(x_n)\rangle$ is a covariant 
function under conformal transformations of $x_i$. Besides, if the dimension $\Delta_i$  reaches the unitary bound  the corresponding operator is conserved 
\bea
\langle \dots \partial_\mu {\mathcal O}^i_{\mu\dots} \dots\rangle=0\ .
\eea
There are other W.I.'s but they are trivially satisfied when all $x_i$ are distinct (which we implicitly assume throughout the paper). 

The covariance under conformal transformations (this also includes Poincar\'{e} symmetry) can be solved in a number of ways, in particular using the embedding formalism \cite{Costa:2011mg}. An explicit expression for the desired correlator will involve a number of  arbitrary functions of the conformal cross-ratios. For the four-point function of the identical operators of dimension $\Delta$ it takes the form
\bea
\label{fdef}
\langle {\mathcal O^1_{\mu_1\dots}}(x_1)\dots  {\mathcal O^4_{\mu_4\dots}}(x_n)\rangle=\prod\limits_{i<j}^4 (x_i-x_j)^{-2\Delta/3}\, \sum_I^N f^I(u,v)\, \IQ_{I\, \mu_1\dots\mu_4\dots}
\eea
The tensor structures $\IQ_I$ are some known expressions made of $x_i^\mu$
and the flat space metric (Kronecker delta-symbol) $\delta_{\mu\nu}$,\footnote{By default we are working in the $\rm d$-dimensional Euclidean space $\R^{\rm d}$, but our results are equally valid in the Minkowski space $\R^{\rm d-1,1}$ after a trivial substitution $\delta_{\mu\nu}\rightarrow \eta_{\mu\nu}$. Another important comment: throughout the paper we focus on the parity-even part of the four-point functions (this does not require the theory to preserve parity). This explains absence of $\epsilon$-tensors inside $\IQ$.} while $u,v$ are defined in \eqref{4pt}. More concretely, each $\IQ$ is a product of certain ``building block'' tensors $H^{(ij)}_{\mu\nu}$ and $V_\mu^{i[jk]}$
and the total number of structures $N$ reflects the number of all possible   combinations of $V$'s and $H$'s resulting in the desired tensor structure of $\IQ$.\footnote{In \eqref{fdef} we slightly altered the definitions of $f^I(u,v)$ (by a factor of $(u^2/v)^{\Delta/3}$) as well as $H$ and $V$ compared with \cite{Costa:2011mg}.} 

Now, in a special case when $\Delta_i$ saturates the unitary bound  (for traceless symmetric operator of spin $\ell$ it is $\Delta={\rm d}+\ell-2$), the derivative $\partial_\mu {\mathcal O}^i_{\mu\dots}$ has the correct property of a primary field of dimension $\Delta+1$. Therefore 
the correlator with ${\mathcal O}^i_{\mu\dots}$ substituted by $\partial_\mu {\mathcal O}^i_{\mu\dots}$ should have a similar representation to \eqref{fdef} albeit with a slightly altered overall prefactor and the new set of tensors $\tilde{\IQ}_{\tilde I}$ and functions $\tilde{f}^{\tilde I}$. The new functions $\tilde{f}$ are related to the original ones through an action of some first order differential operator in variables $u,v$. Thus the conservation of ${\mathcal O}^i$ inside the correlator \eqref{fdef} is equivalent to $\tilde{f}^{\tilde I}=0$ for all $\tilde I$ which can be rewritten in the following way
\bea
\label{ABC}
\left[A^{\tilde I}_I +B^{\tilde I}_I\, {\partial~ \over \partial u}+C^{\tilde I}_I\, {\partial ~\over \partial v}\right]f^I(u,v)=0\ .
\eea
Here $A,B,C(u,v)$ are some rectangular matrices which depend on $x_i^\mu$ only through $u$ and $v$ (these matrices also depend on the dimension $\rm d$ and a choice of the basis for $\IQ$'s). Conservation of each ${\mathcal O}^i$ inside the correlator leads to \eqref{ABC} with its own set of matrices $A,B,C$. With some effort these matrices can be calculated in each particular case (in all cases considered below we calculated $A,B,C$ explicitly using computer algebra). But unfortunately the resulting equations are complicated enough such that we could not find an explicit solution or express it in any other self-contained way. In what follows we will merely analyze these equations with the goal of calculating the number of functional degrees of freedom unconstrained by \eqref{ABC}. We will carry on explaining our logic in a particular case of the correlator of four  conserved currents $J_\mu$ and  return to the four-point function of the stress-energy tensors in the end of this section.

\subsection{Conservation Constraints for Conserved Currents}

In a case of four conserved currents\footnote{For simplicity we assume that all four currents are identical. It is easy to generalize this by introducing a color index $J^a_\mu$, such that each $f^I$ will carry four such indexes $f^{I\, abcd}$.} of dimension $\Delta$ in general $\rm d$ there are $43$ corresponding structures $\IQ$ and $14$ structures $\tilde \IQ$
\bea
\label{jjjj}
\langle J_\mu J_\nu J_\rho J_\sigma\rangle&=&\prod\limits_{i<j}^4 (x_i-x_j)^{-2\Delta/3}\, \sum_I^{43} f^I(u,v)\, \IQ_{I\, \mu\nu\rho\sigma}\ ,\\
\label{ojjj}
\langle {\mathcal O} J_\nu J_\rho J_\sigma\rangle&=&\prod\limits_{i<j}^4 (x_i-x_j)^{-2\Delta/3-\delta_{i,1}/3}\, \sum_{\tilde I}^{14} \tilde f^{\tilde I}(u,v)\, \tilde \IQ_{\tilde I \, \nu\rho\sigma}\ .
\eea 
The scalar operator $\mathcal O$ in \eqref{ojjj} is of dimension $\Delta+1$. In case $\Delta={\rm d}-1$ the conservation condition for the first current $J_\mu$ in \eqref{jjjj} will take the form \eqref{ABC} with the $14\times 43$ matrices $A_1, B_1,C_1$. Similarly, the conservation condition for the second current $J_\nu$ would yield another $14$ equations i.e.~another set of matrices $A_2,B_2,C_2$, etc. All together, there are four conservation conditions, one for each current. Combining them all together we obtain a set of  $56$ equations which can be cast in the form \eqref{ABC} with some $56\times 43$ matrices $A,B,C$. Although there are more equations then unknown functions, these equations are not independent. As a result there are unrestricted degrees of freedom which we wish to identity.

\subsection{Permutation symmetry I}
\label{psi}
Our next step is to reduce the number of functions $f^I$ by imposing the  permutation symmetry which changes the order of operators inside the correlator. In case the operators are bosonic, any permutation should be a symmetry. Otherwise in certain cases the correlator may change sign. For the four point function there are $4!=24$ possible permutations (including the trivial one). But the following $\Z_2\times \Z_2$ subgroup of ${\bf S}_4$ that consists of the following permutations
\bea
\label{spermutations}
(1234)\rightarrow (2143)\ ,\quad (1234)\rightarrow (3412)\ , \quad (1234)\rightarrow (4321)\ .
\eea 
is of particular importance: these permutations leave the cross-ratios $u,v$ invariant.
The action of these permutations on the functions $f^I(u,v)$ is purely algebraic: $f^I(u,v)\rightarrow f^{I}_{\sf new}(u,v)=S^I_J(u,v) f^J(u,v)$. Invariance of $\langle J_\mu J_\nu J_\rho J_\sigma\rangle$ thus reduced to a linear algebra problem of finding the kernel of $(\delta_J^I-S_J^I(u,v))$. We will use the same notations $f^I(u,v)$ to denote vectors from this kernel in some unspecified basis, although now $I$ would run up to $19$ -- the dimension of the kernel in the particular case of four conserved currents. \footnote{Strictly speaking the kernel is parametrized by some new $g^K(u,v)$ with $K=1\dots 19$, such that $f^I(u,v), \  I=1\dots43$, are some linear combinations of $g^K$. But in order to avoid the notation clutter we rename $g^K$ into $f^I$ and hope this will not cause any confusion.}

Invariance with respect to  $\Z_2\times \Z_2$ allows us to reduce the number of the unknown functions from $43$ to $19$. Moreover the number of linearly independent equations reduces to $14$. Indeed, the permutations \eqref{spermutations} are just enough to bring any current out of four to the first position. Hence it is enough to impose conservation of the first current only, all other conservation conditions will be linearly dependent. To conclude, imposing invariance under 
\eqref{spermutations} restricts the problem to the set of equations of the form \eqref{ABC}
with some $14\times 19$ matrices $A,B,C$ and $19$ functions $f^I(u,v)$. In what follows we will refer to these equations as the ``conservation constraints''. 

Let us briefly explain the effect of other permutations. They act on $f^I$ by a combination of linear transformation and a change of variables. For example the permutation 
\bea
(1234)\rightarrow (2134)
\eea
maps $(u,v)$ into $(u',v')\equiv (u/v,1/v)$ and $f^I(u,v)$ into 
\bea
\label{fnew}
f^I_{\sf new}(u,v)=S_{J}^I(u,v)f^J(u',v')\ . 
\eea
The  square matrix $S$ is such that if $f^I(u,v)$ solve the conservation constrains, $f_{\sf new}^I(u,v)$ will also do.  

\subsection{Number of Unrestricted Functional Degrees of Freedom}
Now it is time to return to the conservation constraints \eqref{ABC} and calculate the number of functional degrees of freedom unconstrained by these equations. This can be done using the following simple trick.\footnote{We thank Vasily Pestun for suggesting this idea.} Let us rename the variables $u,v$ into $t,x$ and think of $t$ as ``time'' and $x$ as ``space'' coordinate. Next, we would like to think of \eqref{ABC} as a Cauchy problem,  namely consider \eqref{ABC} as a set of algebraic equations on the ``time'' derivatives ${\partial f^I/ \partial t}$ which we need to express in terms of the original functions $f^I$ and spatial derivatives ${\partial f^I/ \partial x}$. The  $14\times 19$  matrix $B(t,x)$ has rank $12$ which means the conservation constraints \eqref{ABC} can be rewritten as twelve ``time evolution'' equations
\bea
\label{tevol}
{\partial f^i \over \partial t}=F^i\left[f^J, {\partial f^J\over \partial x}\right ]
\eea  for certain twelve $f^i$ (say, $i=1\dots 12$) and two constraints ``without time derivatives''
\bea
\label{Econst}
G_{1,2}\left[f^J, {\partial f^J\over \partial x}\right ]=0\ .
\eea 
It can be checked that the constrains \eqref{Econst} are of the first type. Now one can think of (\ref{tevol}, \ref{Econst}) as of  Cauchy boundary problem for $f^i$ while the remaining $f^{\alpha}$ with $\alpha=13\dots 19$ are  unrestricted and should be thought of as the external parameters.  The ``initial conditions'' are specified by the boundary values $f^i(t^*,x)$ at some boundary $t=t^*$ such that the constraints \eqref{Econst} are satisfied. 

Our first conclusion is that the bose symmetric four-point function of the conserved currents in a general CFT is governed by seven functional degrees of freedom  (we called them $f^\alpha$ above). It is important to note that this number is well-defined and will not change upon a new choice of ``time" direction in the $(u,v)$-plane: the rank of any linear combination of $B$ and $C$ is always $12$.\footnote{This  follows from the explicit form of matrices $B, C$.}

Besides seven unrestricted functions, the correlator also depends on the ``initial conditions'' $f_i(t^*,x)$ at some  $t=t^*$. Furthermore  we still have to impose bose symmetry with respect to the permutations that change $(u,v)$. There are $3!=6$ of those corresponding to the ${\bf S}_3$ group that keeps the first operator inside the correlator in its place and permutes the other three.  We will impose this symmetry in the next subsection.  

\begin{figure}
\begin{tikzpicture}
\draw[step=3.0,white,thin] (0.,0.) grid (12.,8.);    
\node[] at (8.8,1.5) {$w$};
\node[] at (6.5,1.5) {$u$};
\node[] at (.6,7.2) {$v$};  
        \begin{subfigure}[b]{0.5\textwidth}
                \includegraphics[width=0.968\textwidth]{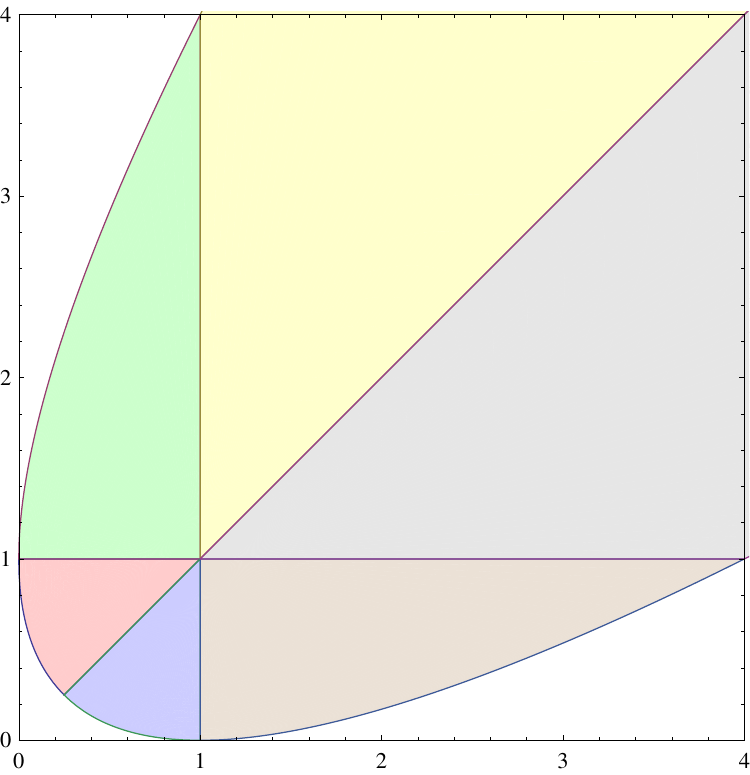}
                \caption{$a).$}
                \label{fig:uv}
        \end{subfigure}%
        ~ 
        \begin{subfigure}[b]{0.5\textwidth}
                \includegraphics[width=1.02\textwidth]{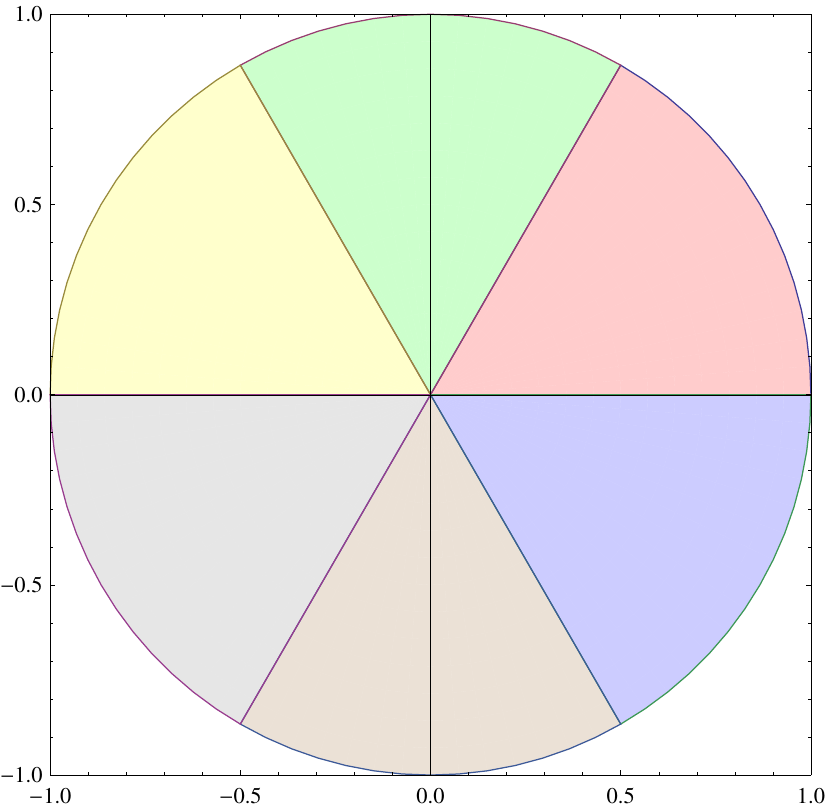}
                \caption{$\quad b).$}
                \label{fig:w}
        \end{subfigure}
\end{tikzpicture}
\caption{(a) The $(u,v)$-plane with physically accessible area of cross-ratios highlighted in color. (b) The same area after a change of variables $(u,v)\rightarrow \omega$.}
\label{fig:plots}
\end{figure}

\subsection{Permutation symmetry II}
Although the equations \eqref{ABC} are defined for any $u$ and $v$, in a Euclidean theory $u$ and $v$ are non-negative, see \eqref{4pt}. In fact the physically accessible points must lie inside the curve $\sqrt{u}+\sqrt{v}=1$ (this area is highlighted in color in Fig.~\ref{fig:uv}). This fact does not invalidate our previous findings based on the Cauchy problem picture on the whole $u, v$ plane (or a quadrangle $u,v\ge 0$) because we can think of the conservation constrains mathematically, without worrying about physical origin and hence scope of $u$ and $v$ (it is important to note that nothing special happens to $A,B,C$ on the boundary  $\sqrt{u}+\sqrt{v}=1$). 

The ``physical" area consists of six patches (each highlighted in its own color in Fig.~\ref{fig:uv}) which are mapped  into each other by ${\bf S}_3$. The permutation symmetry constraints $f^I(u,v)=f_{\sf new}^I(u,v)$  equate $f^I$ from different patches. Thus, at least conceptually,  it is enough to know $f^I$ just in one patch and require that $f^I$ satisfy certain conditions on the patch's boundary (which sometimes is mapped into itself under ${\bf S}_3$) to ensure permutation symmetry. 

Unless the ``time'' and ``space'' coordinates are chosen wisely the permutation symmetry would mix the  unrestricted $f^\alpha$ and dependent $f^i$ degrees of freedom. Therefore the permutation symmetry constraints $f^I(u,v)=f_{\sf new}^I(u,v)$ will necessarily involve all of them. This is definitely not the most concise and desirable way. This complication can be avoided if we choose ``time'' coordinate  coordinate $t$ such that all permutations will map it into itself ${\bf S}_3: (t,x)\rightarrow (t,x')$ where $x'$ is some function of $(t,x)$. In such a case the boundary $t=t^*$ will be mapped into itself, and intuitively we expect the sets of $f^i$'s and $f^\alpha$'s to remain invariant (although $f^i$'s and $f^\alpha$'s would mix between themselves). Indeed, let us rewrite the conservation constraints
in the following form
\bea
\label{cc}
\left[{\mathcal A}^{\tilde I}_J+{\mathcal B}^{\tilde I}_J\, {\partial~\over \partial t}+{\mathcal C}^{\tilde I}_J\, {\partial~\over \partial x}\right]f^J=0\ .
\eea Here we use ${\mathcal A, \mathcal  B, \mathcal  C}$ instead of $A,B,C$ to emphasize that we changed variables $u,v$ into $t,x$ and hence the former are some linar combinations of the latter. For any values of $t,x$ the matrix ${\mathcal B}(t,x)$ has a seven-dimensional kernel which we parametrize by introducing basis elements $\xi_\alpha^I(t,x)$: ${\mathcal B}_J^I\, \xi_\alpha^J=0$. We denote other twelve  linearly independent vectors spanning the space of $f^I(t,x)$  by $\zeta_i^I(t,x)$:
\bea
\label{decomposition}
f^I=f^i \zeta^I_i+f^\alpha \xi_\alpha^I\ .
\eea 
The linear space spanned by $\zeta$'s is not well-defined because one could shift $\zeta$'s by $\xi$'s.  A good way to remove this ambiguity is to introduce a positive-definite metric $g_{IJ}(t,x)$ and require orthogonality of all $\zeta$'s  and  $\xi$'s (i.e.~$\zeta$'s will span the orthogonal complement to the kernel of ${\mathcal B}$). 

Crucially, we will assume that $g_{IJ}(t,x)$ is covariant under the permutation symmetry (now we switch back from $(t,x)$ to $(u,v)$  to stress that covariance of metric  is independent of the choice of variables)
\bea
\label{covmetric}
S^{I}_{I'}(u,v) g_{IJ}(u,v) S^{J}_{J'}(u,v)=g_{I'J'}(u',v')\ .
\eea
Going back to \eqref{decomposition} we define $\zeta$'s such that
$\zeta_i^I(u,v) g_{IJ}(u,v) \xi_\alpha^J(u,v)=0$ for any $i, \alpha$.

In section \ref{psi} we explained that $f^I_{\sf new}$ automatically solves \eqref{cc} so far $f^I$ does. This can only happen if the equations resulting from plugging \eqref{fnew} into \eqref{cc} are linearly dependent with the original equations \eqref{cc} upon a chance of variables $u,v\rightarrow u',v'$. In particular this means (this crucially uses that ``time'' $t$ is mapped into itself:  $t'=t$)
\bea
{\mathcal B}^{\tilde I}(u,v)_{J} S^J_{J'}(u,v)={\bf s}^{\tilde I}_{\tilde I'}(u',v'){\mathcal B}^{\tilde I'}_{J'}(u',v')\ ,
\eea
for some matrix $\bf s$. From here it immediately follows that $S^{I}_J(u,v)\xi_\alpha^{J}(u',v')$ is annihilated by ${\mathcal B}^{\tilde I}_J(u,v)$ and hence it can be expanded in a linear combination of $\xi$'s:
\bea
S^{I}_J(u,v)\xi_\alpha^{J}(u',v')=\xi_\beta^I(u,v) \lambda_\alpha^\beta(u,v)\ ,
\eea
for some matrix $\lambda$. In short, we just derived that the space of $\xi$'s is invariant under  permutations from $\bf S_3$. Since $\zeta$'s were defined as a basis in the orthogonal complement to $\xi$'s and the metric is covariant under $\bf S_3$ we conclude that there is a matrix $w$ such that $S^{I}_J(u,v)\zeta_i^{J}(u',v')=\zeta_j^I(u,v) w_i^j(u,v)$.

Now the permutation symmetry constraint $f^I(u,v)=f_{\sf new}^I(u,v)$ can be rewritten as two separate conditions: one for the unrestricted degrees of freedom $f^\alpha$
\bea
\label{cond1}
\lambda^\alpha_\beta(u,v) f^\beta(u',v') =f^\alpha(u,v)\ ,
\eea
and a similar one for $f^i$'s
\bea
\label{cond2}
w^i_j(u,v) f^j(u',v') =f^i(u,v)\ .
\eea

\subsection{Conformal Bootstrap for Unrestricted D.O.F. }
Conformal bootstrap for a four-point  functions is a combination of two
basic properties of a CFT correlator: crossing symmetry and conformal block decomposition. The latter is just the statement that the corresponding functions $f^I(u,v)$ is a linear combination of some special predetermined functions $G^I_{\Delta, \ell,\dots}(u,v)$ universal for all CFTs in a given dimension $\rm d$, which are called conformal blocks\footnote{Our definition of conformal blocks could differ from the conventional one by a fixed prefactor.} 
\bea
\label{cbd}
f^I(u,v)=\sum_{\Delta, \ell, \dots} c_{\Delta, \ell, \dots} G^I_{\Delta, \ell,\dots}(u,v)\ .
\eea
At the same time crossing symmetry is nothing but the symmetry of the corresponding correlator under permutations of the operators. 
In other words crossing symmetry is summarized in two conditions \eqref{cond1} and \eqref{cond2} derived above. Since $f^i$'s depend on $f^\alpha$'s we suspect the second set of conditions \eqref{cond2}  is redundant. Strictly speaking we can not prove that because besides $f^\alpha$'s functions $f^i(t,x)$ also depend on the boundary conditions $f^i(t^*,x)$. While this is true we believe in a physical theory where $f^I$ satisfy the conformal block decomposition \eqref{cbd} functions $f^\alpha(u,v)$ completely and unambiguously determine $f^i$ and hence \eqref{cond2} follows from \eqref{cond1}. Thus the conformal bootstrap can be conveniently formulated in terms of only unrestricted degrees of freedom
\bea
\label{cbc1}
f^\alpha(u,v)=\sum_{\Delta, \ell, \dots} c_{\Delta, \ell, \dots} G^\alpha_{\Delta, \ell,\dots}(u,v)\ ,\\
\label{cbc2}
f^\alpha(u,v)=\lambda^\alpha_\beta(u,v) f^\beta(u',v')\ .
\eea
Here we introduced $\xi_I^\alpha(u,v)$ such that $\xi_I^\alpha \xi^I_\beta=\delta^\alpha_\beta$,\  $\xi_I^\alpha \zeta^I_i=0$ and 
\bea
f^\alpha=\xi_I^\alpha f^I\ ,\qquad  G^\alpha=\xi_I^\alpha G^I\ .\eea

Certainly our main result (\ref{cbc1}, \ref{cbc2}) would remain just a hypothetical idea unless we can provide the necessary building blocks: permutation covariant positive-definite metric $g_{IJ}$ and a coordinate system $t,x$ such that $t$ is permutation invariant. 
In fact such a  metric  can be easily  constructed by contracting all Lorentz indexes of $Q$'s:
\bea
g_{IJ}(u,v)=Q_{I\, \mu\nu\dots}(x_i) Q_{J\, \mu\nu\dots}(x_i)\ .
\eea

Similarly, finding such $t$ is not very difficult. For example for any function $g(u)$ a sum over all permutations $\sum_{{\bf a}\in {\bf S_3}} g({\bf a}(u))$ or any function of this expression will be permutation invariant, for example
\bea
t=u+v+{u\over v}+{v\over u}+{1\over u}+{1\over v}\ .
\eea
A general $t$ complimented by some $x$ would lead to a very complicated   $\xi$ and $\lambda$ rendering our scheme (\ref{cbc1},\ref{cbc2}) impractical.
To deal with this problem we propose the following coordinate system which incorporates symmetries of the problem in a most natural elegant way.\footnote{We thank Sungjay Lee for help with finding $\omega$.} We rename $t,x$ into $r,\phi$ and introduce $\omega=r\, e^{i\phi}$ through
\bea
u=\left|{1+\gamma\, \omega\over 1\, +\, \omega }\right|^2\ ,\quad 
v=\left|{1+{\bar \gamma}\, \omega\over 1\, +\, \omega }\right|^2\ ,\quad 
\gamma=e^{-2\pi i/3}\ .
\eea
The new variable $\omega$ is related to the canonical variable $z$ defined through $u=z
\bar{z},\ v=(1-z)(1-\bar z)$  through a M$\ddot{\rm o}$bius transform 
\bea
z=(1+\bar\gamma)\, {1+\gamma\, \omega\over 1\, +\, \omega }\ .
\eea
The transformation $(u,v)\rightarrow \omega$ maps the ``physical" area in the $u,v$-plane, Fig.~\ref{fig:uv}, into a ``pizza pie" -- the unit disk on the complex plane split into six equal patches (``slices"), Fig.~\ref{fig:w}. The permutation symmetry $\bf S_3$ acts on $\omega$ ``canonically''  according to a naive geometrical intuition i.e.~``slices" are permuted  while radius $r$ is left intact. The two generators of $\bf S_3$ could be chosen to be 
$\phi\rightarrow -\phi$ and $\phi\rightarrow 2\pi/3 -\phi$. 

\subsection{Small $\rm d$ and Degenerate Tensors}
\label{smalld}
Above we have calculated the number of unrestricted functional degrees of freedom governing the four-point function of conserved currents $\langle J_{\mu}J_\nu J_\rho J_\sigma\rangle$. The calculation (and the result -- seven kinematically-unrestricted functional D.O.F.) was seemingly independent on the dimension $\rm d$. This is not entirely correct: the tensors $\IQ_I$ and $\tilde{\IQ}_{\tilde I}$ from \eqref{jjjj} and \eqref{ojjj} as well as matrices $A,B,C$ depend on $\rm d$. Thus it might happen that some of the properties used in our calculation, such as  rank of $\mathcal B$, are different in some specific dimensions from the generic values. This  does not happen but there is another subtlety associated with small $\rm d$: tensors $\IQ_I$ (and/or $\tilde{\IQ}_{\tilde I}$) could become degenerate. We discuss when exactly this happens in Appendix A. Here we just note that degeneracy of $\IQ_I$ would mean that certain linear combination of $\IQ_I$'s vanish identically for any values of $x_i^\mu$. Correspondingly the metric $g_{IJ}$ is degenerate in this case. This is exactly what happens in $\rm d=3$: two combinations of $\IQ_I$'s are identically zero. Hence two out of seven unrestricted functions are in fact unphysical: they could be chosen to be anything (for example zero) without affecting $\langle J_{\mu}J_\nu J_\rho J_\sigma\rangle$. Hence in $\rm d=3$ the number of unrestricted functional degrees of freedom is in fact five, not seven. 

It was important for us above that while some of $\IQ_I$'s were degenerate, all $\tilde{\IQ}_{\tilde I}$ were linearly independent. Otherwise instead of equating \eqref{ABC} to zero we would have to allow arbitrary functions of $u,v$ along certain directions in $\tilde I$ in the RHS. This would complicate counting the unrestricted degrees of freedom and we avoid discussing such cases.

To conclude, parity-even part of the correlation function of four conserved currents in a general CFT is governed by five in $\rm d=3$ and seven in $\rm d\ge 4$ functional degrees of freedom.\footnote{Throughout this paper we assume $\rm d\ge 3$. The case of $\rm d=2$ would involve too many degeneracies
and is much easier to analyze using different formalism \cite{Belavin:1984vu,CFT}.}  

\subsection{Four Point Function of the Stress-Energy Tensors}
Eventually we are ready to return to the four-point function of the stress-energy tensors. This case is very similar to the four conserved currents discussed above, but the involved matrices are much larger. Thus generically  the four-point function of $T_{\mu\nu}$'s involves 633 independent $\IQ_I$'s:
\bea
\label{tttt}
\langle T_{\mu_1\nu_1} T_{\mu_2\nu_2} T_{\mu_3\nu_3} T_{\mu_4\nu_4} \rangle&=&\prod\limits_{i<j}^4 (x_i-x_j)^{-2\Delta/3}\, \sum_I^{633} f^I(u,v)\, \IQ_{I\, \mu_1\dots \nu_4}\ ,\\
\label{jttt}
\langle J_{\nu_1} T_{\mu_2\nu_2} T_{\mu_3\nu_3} T_{\mu_4\nu_4}\rangle&=&\prod\limits_{i<j}^4 (x_i-x_j)^{-2\Delta/3-\delta_{i,1}/3}\, \sum_{\tilde I}^{302} \tilde f^{\tilde I}(u,v)\, \tilde \IQ_{\tilde I \, \nu_1\dots \nu_4}\ .
\eea 
Here $T_{\mu\nu}$ is an abstract spin two traceless symmetric primary of  dimension $\Delta$ and $J_\nu$ is a primary vector of dimension $\Delta+1$. In reality $\Delta=\rm d$ and $T_{\mu\nu}$ is conserved which results in the equation \eqref{ABC} with $(4\times 302)\times 633$ matrices $A,B,C$. 

Imposing symmetry under permutations \eqref{spermutations} results in the space of $f^I$ being reduced from $633$ to $201$ functions: now the matrices \eqref{ABC} are $302\times 201$. Given that matrix $B$ (or any linear combination of $B$ and $C$) has rank $172$ the number of unrestricted functions is $29$. This a general result valid for $\rm d\ge 6$ when all $\IQ_I$'s and $\tilde \IQ_{\tilde I}$'s are linearly independent. In $\rm d=5$ while all  ${\tilde \IQ}_{\tilde I}$'s are distinct one combination of $\IQ_I$'s vanishes, hence the number of unrestricted functional D.O.F. reduced to $28$. For $\rm d=3,4$ there are many degenerate ${\tilde \IQ}_{\tilde I}$'s (see Appendix A) which complicates further analysis. We will calculate the number of unrestricted functions in this case in the next section using ``duality" between the CFT correlators and scattering amplitudes in an auxiliary Minkowski space.

\section{CFT Correlators and Scattering Amplitudes}
\label{sec:scattampl}
It was first noticed in \cite{Hofman:2008ar} in a particular case of stress-energy tensors and further generalized in \cite{Costa:2011mg} that the number of linearly independent three-point functions of any (conserved or not) primary operators ${\mathcal O}_i$ of spin $\ell_i$ (for simplicity we are talking only about traceless symmetric representations; spin $\ell_i$ is just the number of indexes) in a general $\rm d$-dimensional CFT coincides with the number of linearly independent scattering amplitudes of ``dual'' particles of spin $\ell_i$ in a $\rm d+1$-dimensional Minkowski space. When some operators are conserved, i.e.~when the corresponding $\Delta_i$ saturate the unitary bound, the dual particles should be massless $p_i^2=0$. Otherwise  $p_i^2\neq 0$.

In \cite{Costa:2011mg}  this intriguing coincidence was given the following interpretation. The three-point scattering amplitudes\footnote{In case the involved particles are massless $1\rightarrow 2$ process is prohibited due to kinematics and scattering amplitude in the Minkowski space $\R^{\rm d,1}$ is ill-defined. In such a case the analytic continuaton in momenta into $\R^{\rm d-1,2}$ space is assumed. We thank Jared Kaplan and Leonardo Rastelli for discussing this point.} in a flat space $\R^{\rm d,1}$ can be one-to-one matched   with the cubic interacting vertexes in the Lagrangian. Next, these  vertexes are brought into the $AdS_{\rm d+1}$ space, where through the usual AdS/CFT logic, they give rise to the CFT correlators at the boundary $\R^{\rm d-1,1}$ (or $\R^{\rm d}$ upon a Wick rotation). This picture works well for the three-point functions of any (conserved or not) primaries but its validity and completeness for the four (and higher)-point functions is not clear. Thus the $2\rightarrow 2$ scattering amplitude is not completely determined by the quartic coupling in the Lagrangian, rather it depends on all cubic couplings in the theory. So is the four-point function on the boundary of AdS -- it also depends on all cubic couplings in the bulk. Hence one can envision a matching procedure between the four-point scattering amplitudes in $\R^{\rm d,1}$ and the four-point CFT correlators in $\R^{\rm d}$, but at this point this has not been done. 

Nevertheless it was proved in \cite{Costa:2011mg} that the number of functional degrees of freedom governing $n$-point functions of primary operators ${\mathcal O}_i$ with all $\Delta_i$ above the unitary bound (i.e. no ${\mathcal O}_i$  is conserved) is indeed equal to the number of functions governing scattering amplitudes of $n$ ``dual" massive particles in $\R^{\rm d,1}$. Let us remind the reader how this was established. When all ${\mathcal O}_i$ are non-conserved a general $n$-point function  is given by some generalization of \eqref{fdef} with all functions $f^I$ (which depend on $n(n-3)/2$ conformal cross-ratios) being unconstrained. The corresponding tensors $\IQ_I$'s satisfy certain linear conditions and can be constructed as all possible products of ``building blocks'' $H^{(ij)}_{\mu\nu}$ and $V_\mu^{i[jk]}$ such that the resulting tensors have the desired set of space-time indexes and satisfy necessary symmetries.

Similarly the generic scattering amplitude of $n$ massive particles can be expressed in terms of a sum  \eqref{fdef}
\bea
{\mathcal A}=\sum_{\sf I} {\sf f}^{\sf I}\, \A_{{\sf I}\, M\dots}\ .
\eea
Here ${\sf f}^{\sf I} $ are the functions of $n(n-3)/2$ Mandelstam variables. The tensors $A_{\sf I}$ are build of particle momenta $p^M_i$ and 
Kronecker delta-symbols $\delta_{MN}$\footnote{Let us remind the reader that we focus on the parity-even part of the correlation functions or scattering amplitudes. Relaxing this constraint, i.e.~allowing $\IQ$'s and $\A$'s to include $\epsilon$-tensors would not change the conclusion.}  and satisfy the following equivalence condition (index $M$ corresponds to the $i$-th operator ${\mathcal O}_i$)
\bea
\label{equivc}
{\A}_{\dots M\dots}\simeq {\A}_{\dots M\dots} +p^i_M\, (\dots)\ .
\eea
 It turns out that such $\A_{\sf I}$ can be also constructed as all possible products of  some $H^{(ij)}_{MN}$ and $V_M^{i[jk]}$ (which have the same symmetries as $H^{(ij)}_{\mu\nu}$ and $V_\mu^{i[jk]}$). Since $\A_{\sf I}$'s should have the same index structure as $\IQ$'s we conclude that the spaces of $\IQ_I$'s and $\A_{\sf I}$'s are isomorphic. Although we started with two different sets of linear algebra constraints acting in two different linear spaces, they define isomorphic  linear spaces spanned by $\IQ_I$'s or $\A_{\sf I}$'s. As a result there  are as many $f^I$'s as ${\sf f}^{\sf I}$'s (and that's why we can use $I$ instead of $\sf I$). 

It is only natural now to ask if this relation holds in case when some ${\mathcal O}_i$ are conserved.\footnote{We thank Simone Giombi for posing this question.} Let us stress that in this case we are no longer comparing two linear algebra problems. While the number of independent scattering amplitudes (i.e.~functions ${\sf f}^{\sf I}$) is still governed by linear algebra, the unrestricted $f^I$'s are controlled by the  differential constraint \eqref{ABC}. We could hardly do the comparison in full generality for $n$-point functions, but it is straightforward  to  cover the correlators of four  conserved currents or stress-energy tensors. Indeed, the number of independent $f^I$'s was calculated in the previous section (we covered $\rm d\ge 3$ for conserved currents and $\rm \ge 5$ for stress-energy tensors). The number of linearly independent $\A_{\sf I}$ can be calculated directly using  their definition and properties: the equivalence condition \eqref{equivc} and transversality $p_i^M \A_{\dots M \dots}=0$. Besides, we also impose symmetry with respect to all permutations (this condition should be relaxed in case one is interested  non-bose-symmetric CFT correlators). Remarkably, but not totally unexpectedly,  the results of two calculations perfectly match. Thus we extend and confirm the conjecture of \cite{Hofman:2008ar, Costa:2011mg} to include four-point functions of conserved operators. This gives us a reason to believe the relation between the CFT correlators and scattering amplitudes holds beyond the four-point function, for any combination of primaries, conserved or not.

Strictly speaking, we have only established that the {\it number} of scattering amplitudes matches the {\it number} of CFT correlators. We did not provide any meaningful map between the two spaces. 
But we have little doubt the observed duality is not accidental. Rather it is based on some not yet fully understood physical picture. And therefore such a map must exist, although we expect it to be nontrivial. This is because it will  equate a solution to some {\it linear algebra} problem with a solution to a set of some non-trivial {\it differential} constraints. This prompts us to conjecture that there must be a better formalism to write down the CFT correlators which would not only automatically solve the conformal W.I.'s (like the embedding formalism) but also take care of the conservation constraints (whenever conserved operators are present) reducing them to  linear algebra. Presumably this hypothetical formalism would be the right language to study other properties of CFT correlators, e.g.~impose bootstrap constraints etc.

We summarize our findings in Table \ref{table:1} (for the stress-energy tensors in $\rm d=3,4$ the results are obtained with help of scattering amplitudes only; all other entires are calculated using both approaches: the CFT correlators and the scattering amplitudes). It is quite exciting that a very large original number of $f^I$'s is distilled into a relatively small number of unrestricted functions. Thus in $\rm d=3$ there are just five of those. This strongly  suggests that formulating bootstrap constraints in terms of only unrestricted degrees of freedom is not only a feasible task but also a more practical approach to bootstrap then working with all $f^I$'s.

\begin{table}[]
\centering
\begin{tabular}{| c || c || c | c | c |  } 
\hline
 {\rm correlator} &  {\rm d=3} & {\rm d=4} &  {\rm d=5}  & {\rm d$\ge$ 6}  \\ \hline
\hline
$\langle JJJJ\rangle$  &  5     &    7               &   7   &              7              \\
\hline
$\langle TTTT\rangle$  &  5     &    22               &   28   &              29              \\
\hline
\end{tabular}
\caption{Number of functional degrees of freedom governing bose-symmetric four-point function of conserved currents or stress-energy tensors in a $\rm d$-dimensional CFT.}
\label{table:1}
\end{table}

In conclusion let us briefly discuss yet another method to calculate the number of unrestricted functional degrees of freedom governing the four-point function in an abstract CFT.\footnote{We thank Juan Maldacena and Jo$\tilde{\rm a}$o Penedones for explaining this point to us.} The idea is to use the conformal block decomposition \eqref{cbd} of the correlator \eqref{fdef}. The sum in \eqref{cbd} goes over quantum numbers of the ``intermediate" primary ${\mathcal O}_{\Delta, \ell, {\sf k}}$, i.e.~dimension and Lorentz group representation which we schematically denoted by the total number of space-time indexes  $\ell$ and other quantum number(s) $\sf k$ (in \eqref{cbd} instead of $\sf k$ we simply put dots). The values of  $\Delta, \ell$ are unbounded from above and schematically the sum over two discrete variables $\Delta, \ell$ is responsible for the fact that $f^I$ depend on two variables $u,v$. For each  $\sf k$ and general $\Delta, \ell$ we denote by $n_{\sf k}^1$ the number of linearly independent three-point functions the operator ${\mathcal O}_{\Delta, \ell, {\sf k}}$ can form with ${\mathcal O}^1,{\mathcal O}^2$ and similarly  $n_{\sf k}^2$ for ${\mathcal O}^3,{\mathcal O}^4$. Then the number of independent functional D.O.F. governing the four-point correlator  will be given by the sum $\sum_{\sf k}n_{\sf k}^1\, n_{\sf k}^2$. We illustrate how this formula works in case of four conserved currents in $\rm d=3$ in the \ref{app:B}.

\section{Two Easy Ways to Solve the Conservation Constraints}
We have seen in section \ref{sec:consconst} that imposing conservation in the coordinate space after taking care of  the conformal symmetry leads to a complicated differential constraint. At the same time the intriguing connection with the linear algebra problem of scattering amplitudes discussed in section \ref{sec:scattampl} suggests there should be a better way of solving the whole set of Ward Identities, including conservation of operators. In this section we discuss two straightforward ideas to  explicitly solve the conservation constrains. 

\subsection{Solving All W.I.'s Automatically}
In section \ref{sec:scattampl} we expressed a hope that there should be a mathematically elegant way to solve all Ward Identities including conservation. Indeed, below we present a way to accomplish that \cite{Erdmenger:1996yc}.\footnote{We thank Hugh Osborn for sharing this idea with us.} The price we pay is that not all CFT correlators can be reproduced this way. Hence the problem of finding a better formalism to simultaneously take care of all W.I.'s remains open. 

The key observation is that under certain conditions a derivative of a primary operator is also a primary. Thus for a completely symmetric traceless tensor  with $\ell$ indexes its divergence is a primary when the dimension saturates the  unitary bound $\Delta={\rm d}+\ell-2$. 
Similarly for a completely antisymmetric tensor with $\ell$ indexes its divergence is a primary if $\Delta={\rm d}-\ell$. Say, there is an antisymmetric primary $F_{\mu\nu}$ of dimension $\Delta={\rm d}-2$. Its divergence  $J_\mu=\partial_\nu F_{\mu\nu}$ is a primary vector field of dimension $\Delta={\rm d}-1$. Besides, $J_\mu$ is automatically conserved! 

Let's say we wish to find  the general form of $\langle J_\mu \dots \rangle $, where $J_\mu$ is a conserved current of dimension ${\rm d}-1$ and dots stand for some other primaries. Instead of first solving conformal W.I.'s and then imposing conservation, as we did in section \ref{sec:consconst}, we can use embedding formalism to find the most general form of $\langle F_{\mu\nu} \dots \rangle $ and then simply take a derivative. The result will automatically solve the full set of W.I.'s! It does not matter if there is such an operator $F_{\mu\nu}$ in the CFT in question, or that its dimension may violate the unitary bound. Calculating $\langle F_{\mu\nu} \dots \rangle $ is  just a mathematical trick and prior to taking the derivative it does not correspond to anything physical. Similarly, one can construct the correlators involving the stress-energy tensor starting with the correlators of a fictional primary $C_{([\mu\mu'][\nu\nu'])}$ of dimension $\Delta={\rm d}-2$ which has the symmetries of the Weyl tensor (this is discussed in more detail in \cite{Erdmenger:1996yc}). 

Despite simplicity and obvious advantages, unfortunately not all correlators can be obtained this way. Say, we want to calculate the folllowing three-point function $\langle J_\mu J_\nu J_\rho\rangle$ (to make sure this is non-zero we can further assume the currents carry an extra color index $J_\mu^a$ which we will suppress below). One can readily find there are four linearly independent correlators of this sort, assuming $J_\mu$ is a primary of certain dimension (which we assumed to be $\Delta={\rm d}-1$). After imposing conservation $\partial_\mu J_\mu=0$ only two  combinations survive (we are talking about parity even correlators in a general $\rm d$). Can we reproduce them using the trick with $F_{\mu\nu}$ outlined above? It is easy to show that there are four linearly-independent correlators $\langle F_{\mu\mu'}F_{\nu\nu'}F_{\rho\rho'}\rangle$. But after taking the derivatives all of them become linearly dependent i.e.~there is only one $\langle\partial_{\mu'}F_{\mu\mu'}\, \partial_{\nu'}F_{\nu\nu'}\, \partial_{\rho'}F_{\rho\rho'}\rangle$,
which means the second structure of  $\langle J_\mu J_\nu J_\rho\rangle$ can not be reproduced this way. 

To make the problem even sharper let us consider the two-point function. Conformal symmetry fixes a unique   $\langle F_{\mu\mu'}F_{\nu\nu'}\rangle$, but $\langle\partial_{\mu'}F_{\mu\mu'}\, \partial_{\nu'}F_{\nu\nu'}\rangle$ simply vanishes and can not reproduce the standard two-point function for conserved currents $\langle J_\mu J_\nu \rangle$. We did not find a practical way to describe those correlators that can be obtained using the $F_{\mu\nu}$ trick. Certainly for two and three-point functions this question can be answered by a direct calculation. For a four and higher-point function involving $J_\mu=\partial_\nu F_{\mu\nu}$ this is more complicated. Say, the correlator of the form $\langle F_{\mu\nu}\dots\rangle$ has a decomposition \eqref{fdef} with the functions  $f^I$ which we prefer to denote $f_{F}^{I_F}$. The corresponding correlator $\langle J_\mu\dots\rangle$ will have a similar decomposition parametrized by some other  functions $f^I$ (this time we keep the original notation). Which correlators $\langle J_\mu\dots\rangle$ can be obtained from $\langle \partial_\nu F_{\mu\nu}\dots\rangle$? In terms of $f_{F}^{I_F}$ and  $f^I$ this means there is a first order differential operator 
\bea
\hat{D}^I_{I_F}={\bf A}^I_{I_F}+{\bf B}^I_{I_F} {\partial~ \over \partial u}+{\bf C}^I_{I_F} {\partial~ \over \partial v}\ 
\eea 
which represents taking divergence of $F_{\mu\nu}$.  
It would be interesting to understand which $f^I$'s can be obtained through $f^I=\hat{D}^I_{I_F}f^{I_F}$. In case of many $J_\mu$'s such operators $\hat{D}$ should be combined leading to a differential operator of higher degree. Its nice property is that this operator will be automatically annihilated by the conservation condition \eqref{ABC} for any $f_{F}^{I_F}$'s, but describing the space of possible resulting $f^I$'s is not an easy task.

In case of correlators with the stress-energy tensors obtained through $C_{([\mu\mu'][\nu\nu'])}$ the reason why not all possible structures can be obtained this way is more transparent. The resulting divergence $\partial_\mu T_{\mu\nu}$ is zero identically, even at the coincident points. Hence the resulting correlator $\langle T_{\mu\nu}\dots\rangle$ would not be able to satisfy Ward Identities which include certain contact terms whenever  $\partial_\mu T_{\mu\nu}$ inside the correlator is present. As a result the trick with $C_{([\mu\mu'][\nu\nu'])}$ can only reproduce a part of the answer, as explained in \cite{Erdmenger:1996yc}.

\subsection{Solving Ward Identities in the Momentum Space}
\label{momentumspace}
In section \ref{sec:consconst} we saw that imposing conservation in the coordinate space after taking care of  conformal symmetry led to a complicated problem. What if we invert the order and take care of conservation first and worry about conformal symmetry later? This could be naturally done in the momentum space:  conservation of an operator $\partial_\mu{\mathcal O}_{\mu \dots}=0$ would imply a linear constraint $p^\mu\, {\mathcal P}_{\mu\dots}=0$ for the correlation function $\langle {\mathcal O_{\mu \dots}(p)}\dots\rangle={\mathcal P}_{\mu\dots}$. Such linear constraints can be easily solved explicitly.\footnote{We will see below that in general solving W.I.'s in the momentum space is more challenging than in the coordinate one. Still it has some advantages. This calculation was done for the tree-point function of the stress-energy tensors in $\rm d=3$ in \cite{Maldacena:2011nz}, of scalars in \cite{Coriano:2012wp,Coriano:2013jba}, and more generally of scalars, currents and stress-energy tensors  in \cite{Bzowski:2013sza}.} 

We were a little bit hasty to declare that we would need to solve the homogeneous  constraints $p^\mu\, {\mathcal P}_{\mu\dots}=0$. The Ward Identities responsible for the conservation of ${\mathcal O}_{\mu \dots}$ equate the correlator $\langle \partial_\mu{\mathcal O}_{\mu \dots} \dots\rangle$ not to zero, but to a contact term. Upon taking the Fourier transform the contact term turns into a polynomial in one (or more) of the external momenta (for a simple derivation of Ward Identities in the momentum space see e.g.~\cite{Cappelli:2001pz}). Thus the conservation constraint is taking the form of a system of non-homogeneous linear equations with the known right-hand-side. To illustrate this we turn to the example of the $n$-point function of the conserved currents
\bea
\label{corrJ}
\langle J_{\mu_1}(p_1)\dots J_{\mu_n}(p_n)\rangle ={\mathcal P}_{\mu_1\dots \mu_n}(p_i)\ .
\eea
The conservation constraints then take the form on $n$ equations $p_i^{\mu_i}\, {\mathcal P}_{\dots\mu_i\dots}={\bf P}^i_{\dots \hat{\mu}_i\dots}$ (hat means a skipped index). The right-hand-side ${\bf P}^i$ is some known combination of the $(n-1)$-point functions. Usually one can find  a particular solution of this system explicitly (for example this was done in \cite{Bzowski:2013sza}) or at least this can be done in principle. The main challenge is to find a special homogeneous solution such that the full answer satisfies the conformal Ward Identities. To this end one can write the most general solution of the conservation constraints
\bea
\label{transverseexp}
\langle  J_{\mu_1}(p_1)\dots J_{\mu_n}(p_n)\rangle= \IP^{\sf particular}_{\mu_1\dots \mu_n}(p_i)+\sum_I f^I(p_i\cdot p_j)\,  \IP_{I\,\mu_1\dots \mu_n}\ .
\eea
Here $\IP_I(p_i)$ is the basis in the space of completely transversal  Lorentz-invariant tensors  $p_i^{\mu_i}\, {\IP}_{\dots\mu_i\dots}=0$ made of external momenta $p_i$ and functions $f_I$ depend on all possible Lorentz invariants $p_i\cdot p_j$. 

We deliberately used the same notation for the functions $f^I$ in \eqref{transverseexp} to make it look similar to \eqref{fdef}, although at this point there is not much in common. Indeed  $f^I$'s
from \eqref{fdef} depend on $n(n-3)/2$ conformal cross ratios and $f^I$'s from \eqref{transverseexp} depend on $n(n-1)/2$ Lorentz invariants. Next, the tensor structures $\IQ_I$'s live in the coordinate space and transform covariantly under conformal transformations,  while the tensor structures $\IP_I$'s live in the momentum space and are transversal. Yet, quite unexpectedly the space of $\IQ_I$'s is isomorphic to the space of $\IP_I$'s! Without transversality the tensor structures $\IP$'s are just the general Lorentz-covariant tensor structures $\T_{\mathbb I}$ made  of  $n-1$ external momenta $p_i^\mu$ (here we take into account momentum conservation $\sum^n p_i=0$) and the flat space metric (Kronecker delta-symbol) $\delta_{\mu\nu}$.  As everywhere else in the paper, the $\epsilon$-tensors are excluded because of parity. The tensors $\T_{\mu_1\dots \mu_n}$ can be rewritten as a function $T(z_i)=\T_{\mu_1\dots \mu_n} z_1^{\mu_1}\dots z_n^{\mu_n}$ as is done in \cite{Costa:2011mg}. All  such functions can be built of $H_{(ij)}=z_i\cdot z_j$ and $V_{i[jk]}=z_i\cdot(p_j-p_k)$. This is already very similar to the makeup of $\IQ$'s in the embedding formalism or scattering amplitudes, but at this point there is no constrain that all three indexes $i,j,k$ in $V_{i[jk]}$ must be distinct. Now, we would like to impose transversality. This can be done by multiplying each index by a projector, namely $i$-th index is contracted with a projector $\Pi_i^{\mu_i \tilde\mu_i}=\delta^{\mu_i \tilde\mu_i}-p_{i}^{\mu_i}p_{i}^{\tilde\mu_i}/p_i^2$  (the same method was also used in \cite{Bzowski:2013sza})
\bea
\IP_{\mu_1\dots \mu_n}=\prod_{i=1}^n \Pi_{i\, \mu_i}^{\ \tilde\mu_i}\ \T_{\tilde \mu_1\dots \tilde\mu_n}\ .
\eea
 Clearly, such projectors will annihilate all tensors which include $z_i\cdot p_i$  and therefore the space of linearly independent $V_{i[jk]}$'s should include only those with $i\neq j\neq k$. Besides, $V_{i[jk]}$'s trivially satisfy 
\bea
V_{i[jk]}+V_{i[kl]}+V_{i[lj]}=0\ ,
\eea 
which precisely coincides with the constraint satisfied by $V_{i[jk]}$'s of the embedding formalism, after a trivial redefinition of $V_{i[jk]}$'s. Thus, we have established an isomorphism between the space of $\IP$'s and $\IQ$'s (or scattering amplitudes of massive particles). 

The same logic continue to work if the correlator \eqref{corrJ} also  includes stress-energy tensors or conserved operators of higher spin. For example in case of stress-energy tensor (a conserved traceless symmetric tensor with two indexes) the tensor structures  $\IP_{I\, (\mu_1\nu_1)(\mu_2\nu_2)\dots}$ are not only transversal but also traceless. The corresponding projector then is (see also \cite{Bzowski:2013sza})
\bea\label{ortproj}
\Pi_{\mu\nu\mu'\nu'}=\Pi_{\mu\mu'}\Pi_{\nu\nu'}-{1\over {\rm d}-1} \Pi_{\mu\nu}\Pi_{\mu'\nu'}\ ,
\eea
where $\Pi_{\mu\nu}$ is defined above. Clearly this projector also annihilates all   $V_{i[jk]}$  unless $i\neq j\neq k$, hence establishing isomorphism between $\IP$'s and $\IQ$'s.

We have to note that the argument above is not completely rigorous unless $\rm d$ is large enough such that the dimension of space does not affect the total number of linearly independent tensor structures $\IP$'s  and $\IQ$'s. But for small $\rm d$ there could be degenerate tensors (of the sort discussed in \ref{app:A}) when some combinations of $\IP$'s are zero while their counterparts made of $\IQ$'s are non-trivial (or vice versa). This possible complication can be avoided if we compare $\IP$'s in the $\rm d$-dimensional space with the scattering amplitudes of massive particles $\A$'s in the same space. The scattering amplitudes are equivalent classes 
\bea
\label{sa}
\A_{\dots \mu \dots}\sim \A_{\dots \mu \dots}+p^i_\mu (\dots)
\eea   
in the space of covariant tensors made of $(n-1)$ external momenta $p_i$. 
The tensors $\IP$'s span the linear subspace defined through 
\bea
\label{tr}
p_i^\mu\,  \IP_{\dots \mu \dots}=0
\eea
in the {\it same} space of of covariant tensors made of $(n-1)$ external momenta $p_i$ and $\delta_{\mu\nu}$'s (this is the space spanned by $\T$'s). The isomorphism between the space of the equivalence classes \eqref{sa} and the linear subspace \eqref{tr} is established with help of the  orthogonal projector ($\Pi_{\mu\nu}$ or \eqref{ortproj}) which maps $\A$'s into $\IP$'s. The opposite map is trivial. This method works in any $\rm d$  as it takes care of the  null tensors: a null tensor $\A$ is mapped into  a null $\IP$. Since scattering amplitudes in $\rm d$ dimensions in one-to-one correspondence with the conformal structures in $\rm d-1$ we arrive at the following result: the space of $\IQ$'s in $\rm d$ dimensions is isomorphic to the space of $\IP$'s in $\rm d+1$. 

Now let us return to \eqref{transverseexp} and discuss the conditions on functions  $f^I$ such that \eqref{transverseexp} is conformal. These are the Ward Identities imposing the covariance under dilatation and special conformal transformations. The former is easy to satisfy as it simply requesres $f^I$ to be homogenious functions of certain degree in momenta (it is possible to choose a basis $\IP_I$ such that each element has a definite dimension). The main complexity comes from the special conformal transformations which give rise to a system of second order differential equations. Thus, solving W.I.'s in the momentum space is significantly more involved than in the coordinate space: while the number of unknown functions in both cases $f^I$ is the same (in the momentum space it could be slightly smaller for small $\rm d$), in the momentum space these functions depend on more variables ($n(n-1)/2$ vs.~$n(n-3)/2$) and satisfy a system of second (rather than first) order PDEs. For example, the problem of finding the tree-point function of the stress-energy tensors in the coordinate space is reduced to a simple linear algebra problem (finding a kernel of  $21\times 11$ matrix). In the momentum space the same problem requires solving a bunch of second order PDEs to determine eleven functions of three variables $f^I(p_1^2, p_2^2, p_3^2)$. This was only done recently in \cite{Bzowski:2013sza}.

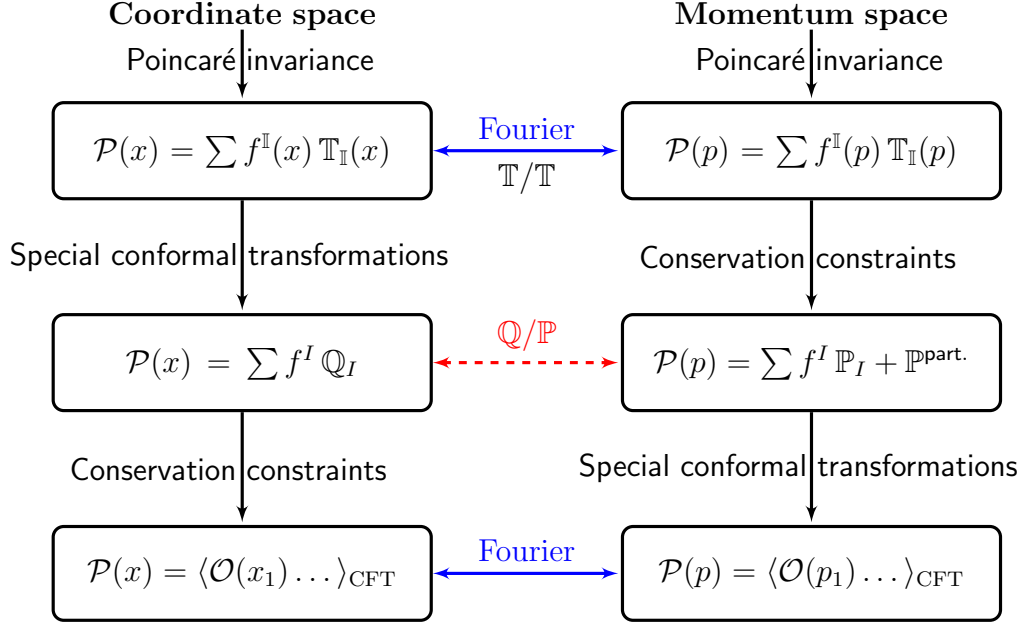
\begin{figure}\begin{center}
\begin{tikzpicture}[node distance=1cm, auto] 
\tikzset{
    mynode/.style={rectangle,rounded corners,draw=black, top color=white, bottom color=white, very thick, inner sep=1em, minimum size=3em, text centered}}
\node[text width=4.2 cm,inner sep=1em, minimum size=3em, text centered] (CS) {\bf Coordinate space};
\node[text width=4.25 cm,inner sep=1em, minimum size=3em, text centered,right=2.5 cm of CS] (MS) {\bf Momentum space};
\node[mynode,text width=4.2 cm, below=.5 cm of CS] (PCS) {${\mathcal P}(x)=\sum f^{\mathbb  I}(x)\, \T_{\mathbb I}(x)$}; 
\node[mynode,text width=4.2 cm, right=2.5 cm of PCS] (PMS) {${\mathcal P}(p)=\sum f^{\mathbb  I}(p)\, \T_{\mathbb I}(p)$};
\node[mynode,text width=4.2 cm, below=1.5 cm of PCS] (CCS) {${\mathcal P}(x)=\sum f^I\, \IQ_I$};
\node[mynode,text width=4.2 cm, right=2.5 cm of CCS] (TMS) {${\mathcal P}(p)=\sum f^I\, \IP_I+\IP^{\sf part.}$};
\node[mynode,text width=4.2 cm, below=1.5 cm of CCS] (SC) {${\mathcal P}(x)=\langle {\mathcal O}(x_1) \dots \rangle_{\rm CFT} $};
\node[mynode,text width=4.2 cm, right=2.5 cm of SC] (SM) {${\mathcal P}(p)=\langle {\mathcal O}(p_1) \dots \rangle_{\rm CFT} $};

\draw[<->, >=latex', color=blue,  shorten >=1pt, very thick]  (PCS.east)  to  node[below] (bottomline) {\textcolor{black}{$\T/\T$}} node[auto] (upperline) {Fourier} (PMS.west);


\draw[<->, >=latex', red, dashed,  shorten >=1pt, very thick]  (CCS.east)  to node[auto] (upperline) {$\IQ/\IP$} (TMS.west);

\draw[<->, >=latex', blue,  shorten >=1pt, very thick]  (SC.east)  to node[auto] (upperline) {Fourier} (SM.west);


\draw[->, >=latex',    shorten >=1pt, shorten <=5pt,very thick] (CS.center)  to node[label={[label distance=-1.7cm]0:{\sf Poincar\'{e} invariance}}] {} (PCS.north);

\draw[->, >=latex',    shorten >=1pt, shorten <=5pt,very thick] (MS.center)  to node[label={[label distance=-1.7cm]0:{\sf Poincar\'{e} invariance}}] {} (PMS.north);

\draw[->, >=latex',    shorten >=1pt, very thick] (PCS.south)  to node[label={[label distance=-3.25cm]0:{\sf Special conformal transformations}}] { } (CCS.north);

\draw[->, >=latex',   shorten >=1pt, very thick] (CCS.south)  to node[label={[label distance=-2.45cm]0:{\sf Conservation constraints}}] {} (SC.north);

\draw[->, >=latex',   shorten >=1pt, very thick] (PMS.south)  to node[label={[label distance=-2.45cm]0:{\sf Conservation constraints}}]{} (TMS.north);

\draw[->, >=latex',   shorten >=1pt, very thick] (TMS.south)  to node[label={[label distance=-3.25cm]0:{\sf Special conformal transformations}}] { } (SM.north);

\end{tikzpicture}
\end{center}
\caption{Parallel between solving W.I.'s in the coordinate and momentum space.}
\label{fig:coordandmoment}
\end{figure}

The observed connection between $\IQ$'s and $\IP$'s suggest that solving W.I.'s in the coordinate and momentum spaces bear in common much more than was realized before. We schematically illustrate this idea in Fig.~\ref{fig:coordandmoment}.
In both coordinate and momentum space one starts by imposing Poincar\'{e} invariance i.e.~representing the correlator of interest ${\mathcal P}_{\mu\dots}\equiv \langle {\mathcal O}_{\mu\dots}\dots\rangle$ as a sum of all possible Lorentz-covariant tensors $\T_{\mathbb I}$'s made of $(n-1)$ linearly independent differences $x_i-x_n$ (or $(n-1)$ linearly independent momenta $p_i$) and Kronecker delta-symbols. The functions $f^{\mathbb I}$'s multiplying $\T$'s depend on all Lorentz-invariant combinations of $x_i-x_n$  or $p_i$. Thus, at this point coordinate and momentum space representations are isomorphic. Another way to establish this isomorphism is through the Fourier transform. 

The final results in the coordinate and momentum spaces, after all W.I.'s are imposed, are obviously related by the Fourier transform as well. What is interesting the intermediate results happens to be related as well. Namely the solution to the special conformal transformations constraints in the coordinate space is related to the solution of the conservation constraints in the momentum space. Strictly speaking this relation (the isomorphism between $\IQ$'s and $\IP$'s) connects the $\rm d+1$-dimensional coordinate space with  the  $\rm d$-dimensional momentum space, but this difference is unimportant for a sufficiently large $\rm d$ (for concrete values see \ref{app:A}). This relation strongly suggests imposing the remaining constraints, covariance under conformal transformations in the momentum space and conservation constraints in the coordinate space, should go in parallel. It is given that the conformal constraints in the momentum space are more comprehensive, as the corresponding functions $f^I$ depend on more variables than their counterparts in the coordinate space. But it should be possible to split the conformal constraints into two groups, such that the first group would reduce the remaining degrees of freedom in $f^I$ to their coordinate space counterpart, while the second group would essentially be equivalent to the analog of \eqref{ABC}.  It would be particularly interesting to try this logic with the three point functions of conserved currents or stress-energy tensors  and explicitly isolate the group of conformal constraints which would be equivalent to the linear algebra constraints imposing conservation in the coordinate space. 

\section{Conclusions}
In this paper we have calculated the number of functional degrees of freedom surviving after imposing the full set of Ward Identities on a four-point function of stress-energy tensors or conserved currents in a $\rm d$-dimensional conformal field theory. The results are presented in Table \ref{table:1}. These numbers precisely match the number of functional degrees of freedom governing the most general scattering amplitude of four gravitons or gauge bosons in $\rm d+1$ dimensions.  Thus our findings support the conjecture that the CFT correlators of primary operators in $\rm d$ dimensions are one-to-one related to the scattering amplitudes of ``dual'' particles in the $\rm d+1$ dimensional space. Quite remarkably this relation connects the linear algebra problem of scattering amplitudes with the differential equations problem of CFT correlators. Hence, we conjecture existence of a new formalism for the CFT correlators which would ``take care'' of all Ward Identities reducing them to a number of linear algebra constraints. 

The number of kinematically unrestricted functions governing the 4pt functions of the stress-energy tensors or conserved current is relatively small, much smaller than the full number of functions before the conservation of operators is taken into account. Therefore we expect that formulating and solving the conformal bootstrap constraints in term of only unrestricted degrees of freedom will have significant advantages over the naive approach which would involve many redundancies. We outlined a way to formulate the conformal bootstrap constraints in terms of only unrestricted degreed of freedom in (\ref{cbc1}, \ref{cbc2}).  

Eventually, we observed an interesting parallel between solving the full set of Ward Identities in the momentum and coordinate spaces. Our findings a schematically illustrated in Fig.~\ref{fig:coordandmoment}.

\section*{Acknowledgments} 
I would like to thank Daniele Dorigoni, Simone Giombi, Sungjay Lee, Juan Maldacena, Hugh Osborn, Miguel Paulos, Jo$\tilde{\rm a}$o Penedones, Vasily Pestun, Slava Rychkov, and Alexander Zhiboedov for discussions. I would also like to thank the Aspen Center for Physics for hospitality and gratefully acknowledge support from a Starting Grant of the European Research Council (ERC STG grant 279617), from the Taplin Fellowship and the grant RFBR 12-01-00482 as well as NSF grant 1066293.

\appendix
\section{Degenerate Tensors in Various Dimensions }\label{app:A}
The main ingredient in our analysis was the linear space of all Lorentz-covariant tensor structures $\T_{\mathbb I}$ made of several vectors $p^\mu_i$ and Kronecker delta-symbols $\delta_{\mu\nu}$. Tensor structures  $\IP_I$ or $\IQ_I$ satisfying transversality  or covariance under special conformal transformations form a subspace in the linear space of all covariant tensors $\T$'s. Naively the space of $\T$'s does not dependent on the dimension of the space $\rm d$. For example for one vector $p^\mu$ there are two linearly independent structures with two indexes 
\bea
T^1_{\mu\nu}(p)=p_\mu p_\nu\ ,\qquad T^2_{\mu\nu}(p)=\delta_{\mu\nu}
\eea
in any $\rm d>1$. But this is not alway the case. Whenever there are several vectors $p_i^\mu$, $1\le i\le n$, certain tensors will  be degenerate (i.e.~identically zero for any values of $p^\mu_i$) for ${\rm d}\le n$. In fact degenerate tensors may appear for $\rm d$ larger than $n$. For example the following tensor is zero in $\rm d=2$  \newpage
\begin{flalign}
\label{2dE}
&\qquad \delta_{\mu\nu}p_{\tilde\mu}p_{\tilde\nu}+p_{\mu}p_{\nu}\delta_{\tilde\mu\tilde\nu}-&
\end{flalign}
\vspace{-1.cm}
\begin{flalign*}
\qquad (p_\mu \delta_{\nu\tilde\mu}p_{\tilde \nu}+p_\nu \delta_{\mu\tilde\mu}p_{\tilde \nu}+p_\mu \delta_{\nu\tilde\nu}p_{\tilde \mu}+p_\nu \delta_{\mu\tilde\nu}p_{\tilde \mu})/2+\\
\end{flalign*}
\vspace{-1.6cm}
\begin{flalign*}
& & p^2(\delta_{\mu\tilde\mu}\delta_{\nu\tilde\nu}+
\delta_{\mu\tilde\nu}\delta_{\nu\tilde\mu}-2\delta_{\mu\nu}\delta_{\tilde\mu\tilde\nu})/2\ .&
\end{flalign*}
This means transversal tensors $\IP_{(\mu\nu)(\tilde\mu\tilde \nu)}(p)$ or conformal tensors $\IQ_{(\mu\nu)(\tilde\mu\tilde \nu)}(p)$ in two dimensions might be degenerate as well (a reader should not be confused by our notations $\IQ(p)$ because $p$ is an abstract vector in $\R^{\rm d}$, not a momentum). We do not know an analytic method to find for which $\rm d$ the degenerate tensors would be present. Therefore we approached this problem empirically and using computer algebra we calculated the scalar product matrix by contracting all Lorentz indexes 
\bea
g_{\mathbb I \mathbb J}(p)=\T_{\mathbb I\, \mu\dots } \T_{\mathbb J\, \mu\dots }
\eea
Thus, we found the scalar product matrix for the tensors $\T_{\mathbb I\, \mu \nu \rho}(p_1,p_2)$ the resulting scalar product is non-degenerate for all $\rm d\ge 3$. Hence all $14$ $\T_{\mathbb I\, \mu \nu \rho}(p_1,p_2)$ are linearly independent and correspondingly all $4$ $\IP_{I\, \mu \nu \rho}(p_1,p_2)$ and $4$ $\IQ_{I\, \mu \nu \rho}(p_1,p_2)$ are non-degenerate. 
Similarly we analyzed the scalar product for $5$ tensors $\T_{\tilde {\mathbb I}\, \mu \nu}(p_1,p_2)$ which also turns out to be non-degenerate for $\rm d\ge 3$ implying linear independence of two $\tilde \IQ_{\tilde I\, \mu \nu}$. 
 Hence one does not have to worry about degenerate tensors while solving Ward Identities for $\langle J_\mu J_\nu J_\rho \rangle$ neither in momentum nor in coordinate space. 

Similarly we analyzed $\T_{\mathbb I\, \mu \nu \rho \sigma }(p_1,p_2,p_3)$ and 
$\T_{\tilde{\mathbb I}\, \mu \nu \rho }(p_1,p_2,p_3)$ ``responsible" for the 4pt function of currents  $\langle J_\mu J_\nu J_\rho J_\sigma \rangle$.  Here $\mathbb I$ runs up to $138$, but in $\rm d=3$ only $81$ and in $\rm d=4$ only $136$ are linearly independent. For $\rm d\ge 5$ there are no degeneracies. Therefore all $43$ transversal tensors $\IP_{I\ ,\mu \nu \rho \sigma }(p_1,p_2,p_3)$ in $\rm d\ge 5$ are distinct, while there are only $41$ of those in $\rm d=4$ and $14$ in $\rm d=3$. Using the isomorphism between $\IP$'s in $\rm d+1$ and $\IQ$'s in $\rm d$ dimensions we conclude that all $43$ conformal structures  $\IQ_{I\ ,\mu \nu \rho \sigma}$ are linearly independent when $\rm d\ge 4$, and there are two degenerate structures in $\rm d=3$. These two degenerate structures are responsible for the difference between the number of unrestricted functions governing $\langle J_\mu J_\nu J_\rho J_\sigma \rangle$ in $\rm d=3$ and all other dimensions $\rm d\ge 4$ (see section \ref{smalld}). It is important to note that all $14$ $\tilde \IQ_{\tilde I\,\mu \nu \rho}$ in $\rm d\ge 3$ are independent and therefore the analysis of section \ref{sec:consconst} in coordinate space is valid. At the same time not all $36$ $\tilde{\T}_{\tilde{\mathbb I}\,\mu \nu \rho}(p_1,p_2,p_3)$ are independent in $\rm d=3$, in fact there are $9$ degenerate tensors of this kind. That is why solving Ward Identities for $\langle J_\mu J_\nu J_\rho J_\sigma \rangle$ in $\rm d=3$ in the momentum space would require extra care: the RHS of the conservation constraint ${\bf P}^i$ introduced  the section \ref{momentumspace} may include extra terms which is just zero in disguise.  

Before we turn to discussing the tensor structures relevant for correlators of the stress-energy tensors let us explain the origins of \eqref{2dE}.  Let's introduce an auxiliary metric $g_{\mu\nu}(x)$ on an asymptotically flat $\R^2$ space. The functional $W[g_{\mu\nu}]=\int\hspace{-.1cm}\sqrt{g}\, R$ in $\rm d=2$ is trivial -- it calculates Euler characteristic which is a  topological quantity and hence does not depend on $g_{\mu\nu}(x)$. Thus, the first variational derivative ${\delta W/ \delta g^{\mu\nu}(x)}$ calculated in a flat space $g_{\mu\nu}=\delta_{\mu\nu}$ is just zero.  Yet the second derivative  
\bea
\left.{\delta^2 W\over \delta g^{\mu\nu}(x)\, \delta g^{\tilde\mu\tilde\nu}(y)}\right|_{g_{\mu\nu}=\delta_{\mu\nu}}
\eea
will give a non-trivial expression which is zero in disguise. Upon the Fourier transform with respect to $x-y$ one obtains 
\eqref{2dE} which explains why it is degenerate in $\rm d=2$. The Euler characteristic exists in any even-dimensional space $\rm d=2m$. Written in terms of local metric it is proportional to $m$-th power of Riemann curvature. Therefore first $\rm m$ variational derivatives with respect to metric will vanish identically, while the $\rm m+1$-th (and all higher derivatives) upon the Fourier transform would lead to a degenerate tensor in a $\rm d=2m$-dimensional space. We observed that it would be the only degenerate tensor $\T_{(\mu_1\nu_1)\dots (\mu_{\rm m+1}\nu_{\rm m+1})}(p_1,\dots, p_{\rm m})$ with $\rm m+1$ symmetric pairs of indexes depending on $\rm m$ independent vectors in $\R^{2\rm m}$. Moreover there would be no degenerate tensors of this kind in $\rm d>{2\rm m}$ and several (or many) in $\rm d<{2\rm m}$. Furthermore there are no degenerate tensors with one index less $\T_{\mu_1(\mu_2\nu_2)\dots (\mu_{\rm m+1}\nu_{\rm m+1})}(p_1,\dots, p_{\rm m})$ in $\rm d\ge {2\rm m}$
and several (or many)  $\rm d<{2\rm m}$.

This simple observation can help up understand when one has to worry about degenerate tensor structures while dealing with the $n$-point function of the stress-energy tensors. Let us start with the 2pt function. The corresponding general Lorentz-covariant tensors $\T_{\mathbb I\, (\mu\nu)(\tilde \mu\tilde \nu)}(p)$ depend on one vector $p^\mu$ and are symmetric with respect to $\mu\leftrightarrow \nu$ and  $\tilde\mu\leftrightarrow \tilde\nu$. We do not require  $\T_{\mathbb I}$'s to be traceless. There are $6$ of those overall and as we discussed above exactly one becomes degenerate in $\rm d=2$. We have already mentioned that all tensor structures $\tilde \T_{\tilde{\mathbb I}\, 
\mu(\tilde \mu\tilde \nu)}(p)$ are non-degenerate in $\rm d\ge 2$ (and so obviously would be any tensor with less number of indexes). Therefore the unique  zero structure in $\rm d=2$ is traceless and transversal i.e.~it is one of those which we called $\IP_I$ above. In fact there is just one traceless transversal tensor $\IP_{(\mu\nu)(\tilde \mu\tilde \nu)}(p)$ in any $\rm d$ and the fact that it is degenerate in $\rm d=2$ is directly responsible for the conformal anomaly: as soon as $T_{\mu\nu}$ is conserved the 2pt function $\langle T_{\mu\mu} (p) T_{\nu\nu} (-p) \rangle$ can not be zero.  Since $\IQ$'s in $\rm d$ dimensions are isomorphic to $\IP$'s in $\rm d+1$,  nothing pathological happens with the unique conformal structure  $\IQ_{(\mu\nu)(\tilde \mu\tilde \nu)}(x-y)$ in $\rm d\ge 2$. Hence the 2pt of the stress-energy tensors in coordinate space is uniquely fixed and is well-defined in all $\rm d$.

Similarly, there is no degenerate tensor structures with three symmetric pairs of indexes that depend on $p_1,p_2$ in $\rm d>4$ (there are $137$ of them in total). In $\rm d=4$ exactly one traceless transverse tensor becomes zero. There are already $25$ zero $\T_\mathbb I$'s in $\rm d=3$. Therefore the number of transverse traceless $\IP$'s drops from $11$ in $\rm d>4$ to $10$ in $\rm d=4$ and $4$ in $\rm d=3$. These degenerate tensor in $\rm d=3, 4$ should be taken into account while solving the Ward Identities in the momentum space \cite{Cappelli:2001pz,Bzowski:2013sza}. Eventually, the degenerate tensor $\IP$ in $\rm d=4$ implies there is exactly one degenerate conformal structure $\IQ$ in $\rm d=3$. And that is why there is one less linearly independent parity-even 3pt functions of the stress-energy tensors in $\rm d=3$: $2$ instead of $3$.

Finally, let us discuss the 4pt function of $T_{\mu\nu}$'s. In $\rm d\ge 6$ there is no degenerate $\IQ$'s but there is exactly one zero $\IQ$ (out of $633$) in $\rm d=5$ (which is ``dual'' to the unique degenerate $\IP$ in $\rm d=6$). This degenerate conformal tensor  is the reason why the number of  unrestricted functional D.O.F. governing the 4pt in $\rm d=5$ is by one smaller than in $\rm d=6$ (see Table \ref{table:1}). 

We conclude this section with a technical note. To calculate the scalar product matrix $g_{IJ}$ for the $633$ conformal structures $\IQ_{I\, (\mu_1\nu_1)\dots (\mu_4\nu_4)}(p_1,p_2,p_3)$ discussed above is not quite trivial because it requires first calculating a larger matrix $g_{\mathbb I \mathbb J}$ for the most general  tensors $\T_{\mathbb I\, (\mu_1\nu_1)\dots (\mu_4\nu_4)}(p_1,p_2,p_3)$. The problem is that $g_{\mathbb I \mathbb J}$ is rather large, namely $6536\times 6536$. Thus it would be desirable to find a way to calcualte the scalar product of $\IQ$'s directly, without defining it in the larger space of $\T$'s first. This is in fact easy to do for the tensor structures $\IQ$ which correspond to any four-point function i.e.~which depend on three vectors. The tensor structure of $\IQ$'s i.e.~spin of corresponding operators is unimportant.  Indeed, using conformal symmetry one can bring four vectors $x_i^\mu$ to a ``canonical'' form when the first one vanishes, the fourth is at infinity, and the third is at unite distance along some direction $\vec{e}_1$. The remaining freedom is the location of the second point in a plane spanned by two vectors $\vec{e}_1, \vec{e}_2$: $\vec{x}_2=a\, \vec{e}_1+ b\,\vec{e}_2 $. Hence the space of conformal structures $\IQ_I (p_1,p_2,p_3)$ is isomorphic to the space of all covariant tensors which depend on two vectors $p_1=a\, \vec{e}_1$ and $p_2=a\, \vec{e}_2$. Since in case of the stress-energy tensors we require $\IQ$'s to be traceless, we should impose this condition on $\T$'s as well. Finally we obtain that the space of $633$ conformal structures $\IQ_{I\, (\mu_1\nu_1)\dots (\mu_4\nu_4)}(p_1,p_2,p_3)$ is isomprohic to the space of general traceless tensors $\T^{\sf traceless}_{\mathbb I\, (\mu_1\nu_1)\dots (\mu_4\nu_4)}(p_1,p_2)$. Similarly the $43$ conformal structures $\IQ_{I\, \mu\nu\rho\sigma }(p_1,p_2,p_3)$ corresponding to the four-point function of currents are isomorphic to $\T_{\mathbb I\, \mu\nu\rho\sigma }(p_1,p_2)$.

\section{Conformal Block Decomposition in $\rm d=3$}\label{app:B}
In this section we will calculate the number of unrestricted functional degrees of freedom governing the four-point function of the conserved currents in $\rm d=3$ using the conformal block decomposition.\footnote{This calculation was done together with Jo$\tilde{\rm a}$o Penedones.}  The general idea was explained in the end of section \ref{sec:scattampl}. Since in $\rm d=3$ dimensions the only possible operators are traceless symmetric tensors with $\ell$ indexes the number of unrestricted degrees of freedom governing a correlator of four conserved currents is given by $n^2$ where $n$ is a number of three-point functions $\langle J J {\mathcal O}_{\mu_1\dots \mu_{\ell}} \rangle$. This number was found in \cite{Costa:2011dw} to be $4$. Hence there are $4^2=16$ functional degrees of freedom governing $\langle JJJJ \rangle$ in $\rm d=3$. Since $2$ out of $4$ three-point functions $\langle J J {\mathcal O}_{\mu_1\dots \mu_{\ell}} \rangle$ are parity-even and the other two are parity odd the $16$ functions split into $8+8$ responsible for the parity-even and parity-odd parts of $\langle JJJJ \rangle$ correspondingly. Now we would like to analyze the action of the permutation group $\Z_2\times \Z_2$ \eqref{spermutations}. Depending on $\ell$ the two parity-even three-point functions are both symmetric or antisymmetric under the exchange of two $J$'s. The resulting parity-even $4$ functions contribute to the completely  $\Z_2\times \Z_2$-symmetric part of $\langle JJJJ \rangle$. Two parity-odd three-point functions have opposite symmetry with respect to permutation of $J$'s for the given $\ell$. Hence the corresponding $4$ functions split as follows: one is completely $\Z_2\times \Z_2$-symmetric, while the other three are odd with respect to two of the three generators \eqref{spermutations} and invariant with respect to the remaining one. Eventually we have $5$ functions governing the parity-even  $\Z_2\times \Z_2$-symmetric part of $\langle JJJJ \rangle$ and three functions each governing a non-trivial representation of $\Z_2\times \Z_2$. This counting matches our findings from section \ref{smalld} (including non-trivial representations of  $\Z_2\times \Z_2$, although we did not mention them there explicitly).


\end{document}